\documentclass[12pt,a4paper]{article}
\usepackage{a4wide}
\usepackage[centertags]{amsmath}
\usepackage{amssymb}
\usepackage{epsfig}
\usepackage{cite}
\usepackage{ulem}
\usepackage{float}
\usepackage[T1]{fontenc}
\usepackage{textcomp}

\newcommand{\drawsquare}[2]{\hbox{%
\rule{#2pt}{#1pt}\hskip-#2pt
\rule{#1pt}{#2pt}\hskip-#1pt
\rule[#1pt]{#1pt}{#2pt}}\rule[#1pt]{#2pt}{#2pt}\hskip-#2pt
\rule{#2pt}{#1pt}}
\newcommand{\fund}{\raisebox{-.5pt}{\drawsquare{6.5}{0.4}}}
\newcommand{\Ysymm}{\raisebox{-.5pt}{\drawsquare{6.5}{0.4}}\hskip-0.4pt%
         \raisebox{-.5pt}{\drawsquare{6.5}{0.4}}}
\newcommand{\Yasymm}{\raisebox{-3.5pt}{\drawsquare{6.5}{0.4}}\hskip-6.9pt%
        \raisebox{3pt}{\drawsquare{6.5}{0.4}}}
\newcommand{\antifund}{\overline{\fund}}

\renewcommand{\thefootnote}{\fnsymbol{footnote}}

\def\be{\begin{equation}}
\def\ee{\end{equation}}
\def\bea{\begin{eqnarray}}
\def\eea{\end{eqnarray}}

\begin{document}

\thispagestyle{empty}

\begin{flushright}
  IPPP/08/43\\
  DCPT/08/86\\
\end{flushright}
\vskip 2cm

\begin{center} {\Large \bf Oddness from Rigidness}
\end{center}
\vspace*{5mm} \noindent \vskip 0.5cm \centerline{ \large Stefan
  F\"orste$^{a}$\footnote{ E-mail address: 
    forste@th.physik.uni-bonn.de} and Ivonne Zavala$^{b}$\footnote{
    E-mail address: Ivonne.Zavala@durham.ac.uk}}

\vskip 1cm \centerline{$^{a}$ \it Physikalisches Institut,
  Universit\"at Bonn} \centerline{\it Nussallee 12, D-53115 Bonn,
  Germany} \vskip0.5cm \centerline{$^{b}$ \it Institute for Particle
  Physics Phenomenology (IPPP)} \centerline{\it South Road, Durham DH1
  3LE, United Kingdom} \vskip2cm

\centerline{\bf Abstract} \vskip .3cm We revisit the problem of
constructing type IIA orientifolds on $T^6/{\mathbb Z}_2 \times
{\mathbb Z}_2$ which admit (non)-factorisable lattices. More concretely,
we consider a ${\mathbb Z}_2 \times {\mathbb Z}_2'$ orientifold with torsion, where
D6-branes wrap {\it rigid} 3-cycles.  We derive the model building
rules and consistency conditions in the case where the
compactification lattice is non-factorisable. We
show that in this class of configurations, (semi) realistic models
with an {\it odd} number of families can be easily constructed, in contrast
to compactifications where the D6-branes wrap non-rigid cycles. We also show that an odd number of families can be obtained in
the factorisable case, without the need of tilted tori.  We illustrate
the discussion by presenting three family Pati-Salam models with no
chiral exotics in both factorisable and non-factorisable toroidal compactifications.

\vskip .3cm

\newpage
%
%
\renewcommand{\thefootnote}{\arabic{footnote}}
\setcounter{footnote}{0}
\section{Introduction}

Type II string theory orientifold compactifications can lead to
effective four dimensional theories with gauge symmetries, chiral
spectrum of fermions and ${\cal N}=1$ supersymmetry. Hence, they
constitute a candidate string theory incorporating real particle
physics. 
In particular, type IIA  toroidal orientifolds with intersecting
D-branes at angles have become extremely popular in the last years
\cite{reviews},  due in part to their relative simplicity and thus
calculability.  
Recent developments on these  models aim to provide more
realistic scenarios, as well as a better understanding of these
constructions.\footnote{Meanwhile heterotic
  orbifold constructions have  been improved towards realistic particle
  physics
  \cite{Kobayashi:2004ud,Forste:2004ie,Buchmuller:2004hv,Lebedev:2006kn}. 
 }  

A recent  development was achieved in \cite{bcms} (see also \cite{ms,dt})
where the authors 
considered a type IIA string theory compactified on a factorisable
$T^6/{\mathbb Z}_2\times{\mathbb Z}'_2$ orientifold with {\it torsion}
\cite{torsion}. This type of construction admits collapsed or {\it
  rigid 3-cycles}, where intersecting D6-branes can wrap. Thus such
D-branes cannot leave orbifold fixed points. 
This fact permitted the authors of \cite{bcms} to
build chiral intersecting D6-brane models with (almost) absent
open string moduli. In other words, massless adjoint fields
associated to the D6-brane 
positions can be  removed from the spectrum, and asymptotic freedom is
easier to achieve. 
However, the models studied in \cite{bcms}, 
consist of  four families, which makes them  phenomenologically unattractive.

An interesting generalisation to the standard factorisable IIA
orientifolds  usually considered in  
the literature \cite{reviews} was performed in \cite{bcs,ftz,kot},
where more general compactification 
lattices were allowed.  In particular in \cite{ftz} non-factorisable 
$T^6/{\mathbb Z}_2\times{\mathbb Z}_2$ orientifolds (without torsion)
were studied. In that paper, D6-brane configurations giving rise to
chiral matter on the 4D spacetime were investigated. It was
found  that intersecting D6-brane models with non-factorisable
compactification lattices, give always rise to even number of
families. This observation resulted in unrealistic particle physics models, thus
disfavoured in comparison with their factorisable cousins. 

It is thus natural to ask  whether the unsatisfactory  phenomenological result found in
\cite{ftz}, can be overcome in 
compactifications which admit non-factorisable lattices in addition to 
rigid cycles where D6-branes can wrap. This is the main question we
investigate in the present paper.   

We find that once rigid cycles are present, it is possible to obtain an odd number of families, as opposed to non-factorisable orientifold models without torsion. Model building rules in this compactifications  depend on the non-factorisable lattice, just as in the case studied in  \cite{ftz} for the $T^6/{\mathbb Z}_2\times{\mathbb Z}_2$ orientifold (without torison).  
Encouraged by these observations, we illustrate the model building rules explicitly by
constructing a {\it three family} Pati-Salam model\footnote{We focus
  on the Pati-Salam instead of the Standard Model gauge group in order
  to automatically satisfy K-theory constrains \cite{bcms}. This
  implies that we consider always an even number of D6-branes
  per stack, which then implies a gauge group U$(2N)$.}. The model
preserves ${\cal N} =1$ supersymmetry and contains
the chiral spectrum of a three family Pati-Salam model. Mass terms for
all additional fields can be written down without breaking the
Pati-Salam gauge group, i.e.\ there are no chiral exotics.

We  go beyond our original motivation and reconsider factorisable 
lattices of $T^6/{\mathbb Z}_2\times{\mathbb Z}'_2$ (with
torsion). Following the same 
strategy as in the non-factorisable case, we find that factorisable
lattices too admit an odd number of families.  Furthermore, this has the bonus that
no tilted tori are required, as it is the case with non-rigid
factorisable models \cite{csu,cll}.  Thus we succeed in providing
examples of three-family models in factorisable and 
non-factorisable lattices with rigid branes, on toroidal
orientifold compactifications  with torsion.

The paper is organised as follows. In the next section we discuss in a
general setup the properties of orientifold constructions with rigid
cycles valid for factorisable and non-factorisable lattices. We
present tadpole constraints, spectrum and supersymmetry conditions.  In
section 3 we illustrate the details of the construction in a fully
worked out example. We first look at a non-factorisable supersymmetric
${\mathcal N}=1$ three-family model using {\it rigid} visible sector branes
as well as hidden {\it semi-rigid} and {\it non-rigid} branes, as will be  explained in the text. We discuss tadpoles, spectrum
and supersymmetry conditions. We then present the factorisable version
of this model, showing how odd number of families can be obtained from
rigid branes without the need of introducing tilted tori. We
close in section 4 with our conclusions. 

Throughout the paper, we  make extensive use of the results of
\cite{bcms} and \cite{ftz}, which we advise the reader to consult for
more details.

\section{Orientifolds with rigid cycles}

In this section we describe the procedure and rules to construct
intersecting D6-brane models on $T^6/{\mathbb Z}_2\times{\mathbb
  Z}_2'$ orbifolds with discrete torsion \cite{torsion,bcms}.

Consider type IIA theory compactified on $T^6/{\mathbb
  Z}_2\times{\mathbb Z}_2'$ where the ${\mathbb Z}_2$ generators act
as \be \theta: z^{1,2} \to -z^{1,2} \,, \qquad \quad \theta^\prime :
z^{2,3} \to -z^{2,3}
\label{eq:orbiac}
\ee
on the three complex coordinates of the compact space.  

Extending the discussion of \cite{bcms}, we allow the $T^6$ lattice to
be either factorisable or non-factorisable, i.e.\ the factorisation $T^6 = (T^2)^3$ is not
respected by the orbifold action.  Moreover, we choose our
compactification  such that fundamental lattice vectors can be
expressed as integer linear combinations of fundamental vectors in the
factorisable lattice. The factorisable lattice is a product of three
$T^2$ lattices where each $T^2$ is obtained by compactification of the
complex planes spanned by the coordinates appearing in
(\ref{eq:orbiac}). The fundamental cycles on these $T^2$ are denoted
by $\left[ a^i\right]$ and $\left[ b^i\right]$, $i=1,2,3$,
\begin{equation}
  \begin{aligned}
    \left[ a^1 \right] = \left( 1,0,0,0,0,0\right) ,\hspace*{1in} &
    \left[ b^1 \right] = \left( 0,1,0,0,0,0\right) , \\
    \left[ a^2 \right] = \left( 0,0,1,0,0,0\right) ,\hspace*{1in} &
    \left[ b^2 \right] = \left( 0,0,0,1,0,0\right) , \\
    \left[ a^3 \right] = \left( 0,0,0,0,1,0\right) ,\hspace*{1in} &
    \left[ b^3 \right] = \left( 0,0,0,0,0,1\right) ,
  \end{aligned}
  \label{eq:halfyc}
\end{equation}
in real coordinates, $x^a$, $a=1, \ldots, 6$, which are related to the
complex coordinates in (\ref{eq:orbiac}) as $z^I = x^{2I-1} +
\mbox{i}\, x^{2I}$, $I=1,2,3$.

It is convenient to give wrapping numbers always with respect to the
factorisable basis as we do in the rest of the paper. This implies
that on non-factorisable lattices not all integer wrapping numbers are
allowed (see \cite{ftz}).

\subsection{Rigid cycles}

Let us consider first the covering space $T^6$. We introduce D6-branes
at angles, which are specified by wrapping numbers $(n^i,m^i)$ along
$[a^i]$ and $[b^i]$. Thus an orbifold invariant D6-brane labelled $a$
wraps the three-cycle: 
\be \label{3-cycle}
\Pi_a^{T^6} = \bigotimes_{i=1}^3 \left( n^i_a\
  [a^i] + m^i_a\ [b^i] \right) .
\ee
As explained in \cite{bcms}, these cycles of $T^6$ are inherited
by the orbifold quotient.  Moreover under the action of
$\mathbb{Z}_2\times \mathbb{Z}_2'$, a three-cycle on $T^6$ has three
images, all of them with the same wrapping numbers as the initial
three-cycle. Therefore, a three-cycle can be identified with
$[\Pi_a^B] = 4\, [\Pi_a^{T^6}]$. Computing the intersection number of
two such cycles gives
\be \label{prod} 
[\Pi_a^B] \cdot [\Pi_b^B] = 4\, [\Pi_a^{T^6}] \cdot
[\Pi_b^{T^6}] 
\ee
where $ [\Pi_a^{T^6}] \cdot [\Pi_b^{T^6}]$ has to be worked out for
each non-factorisable lattice separately as was shown in \cite{ftz}
(see also section \ref{example}).

Besides these untwisted cycles there are also independent collapsed
three-cycles for each of the three twisted sectors, $\theta$,
$\theta'$ and $\theta\theta'$.  In order to determine these, we need
to know the fixed points associated to the compactification
lattice. For non-factorisable tori, these
have to be found in each lattice independently. We perform this
counting explicitly in the next section. Here we give general
expressions for a given lattice.
 
Consider first the $\theta$ twisted sector. We denote the location of
the fixed torus on the first two complex planes by $[E^\theta_{I_a}]$,
where $I_a$ labels the fixed point through which a stack of branes
$D_a$ passes in this sector.  For the ${\mathbb Z}_2 \times {\mathbb
  Z}_2 ^\prime$ orbifold these fixed points correspond to collapsed
two-cycles in the blown up Calabi--Yau space.  These two-cycles are
combined with a one-cycle in the third plane $n^3 [\tilde{a}^3] + m^3
[\tilde{b}^3]$ in order to form a three-cycle in the $\theta$-twisted
sector. Here, $[\tilde{a}^3]$ and $[\tilde{b}^3]$ generate the
$\theta$-fixed torus. For the factorisable lattice, they coincide with
$\left[a^3\right]$ and $\left[b^3\right]$. Let us denote a basis of
such twisted three-cycles as
\be\label{twistedbase} [\alpha^\theta_{I,\,n}] = 2\,
[E^\theta_{I}]\otimes [\tilde{a}^3], \quad\quad
[\alpha^\theta_{I,\,m}] = 2\, [E^\theta_{I}]\otimes [\tilde{b}^3]\,.
\ee
The extra factor of two is due to the action of $\theta'$ on the
twisted three-cycles in the third complex plane. Analogously, the
basic twisted three-cycles in the $\theta'$ and $\theta\theta'$
twisted sectors are defined as
\be
\begin{array}{lcl} \vspace*{.2cm} \quad [\alpha^{\theta'}_{I,\,n}] =
  2\, [E^{\theta'}_{I}]\otimes [\tilde{a}^1], & &
  [\alpha^{\theta'}_{I,\,m}] = 2\, [E^{\theta'}_{I}]\otimes [\tilde{b}^1], \\
  \quad [\alpha^{\theta\theta'}_{I,\,n}] = 2\,
  [E^{\theta\theta'}_{I}]\otimes [\tilde{a}^2], & &
  [\alpha^{\theta\theta'}_{I,\,m}] = 2\, [E^{\theta\theta'}_{I}]
  \otimes [\tilde{b}^2].
\end{array}
\ee

\noindent The intersection number between a pair of such cycles is easy to
compute knowing that $[E_{I}^g] \cdot [E_{J}^h] = -2 \delta_{IJ}\,\delta^{\,gh}$.  Thus the
full twisted three-cycles are given by
\be\label{3cycle} 
[\Pi^g_{I,a}]= n_a^{i_g} [\alpha^g_{I,\,n}] +
m_a^{i_g} [\alpha^g_{I,\,m}] .  
\ee
Given two three-cycles
$$[\Pi^g_{I,a}]= n_a^{i_g} [\alpha^g_{I,\,n}] 
+ m_a^{i_g} [\alpha^g_{I,\,m}]$$ and
$$[\Pi^h_{J,b}]= n_b^{i_h}[\alpha^h_{J,\,n}] 
+ m_b^{i_h} [\alpha^h_{J,\,m}]\,,$$ with $g,h = \theta, \theta',
\theta\theta'$, the intersection between them is 
\be \label{fractional1} 
[\Pi^g_{I,a}] \cdot [\Pi^h_{J,b}]\, =\, 4\,
\delta_{IJ} \delta^{\,gh} \, (n_a^{i_g}\, m_b^{i_g} - m_a^{i_g}\,
n_b^{i_g})\, =\, 4\, \delta_{IJ} \delta^{\,gh} \, (n_a^{i_g}\, m_b^{i_g}
- m_a^{i_g}\, n_b^{i_g}) , 
\ee
where we have again identified intersection points under the orbifold
action and we have used that $[\tilde a^i] \cdot [\tilde b^j] = -
\delta_{ij}$. In this notation, for the twisted sectors $g = \theta,
\theta',\theta\theta'$ one has $i_g = 3,1,2$, respectively.

Now that we know how to describe the non-factorisable untwisted and
twisted sector three-cycles, we construct rigid D6-branes in this 
setup. That is, we consider fractional D6-branes which are wrapping
special Lagrangian 3-cycles, and are charged under all three different
twisted sectors of the orbifold.  The location of a fractional
D6-brane has to be invariant under the orbifold action and thus it
must run through four fixed points for each twisted sector.  Denoting
this set of fixed points as $S^a_g$, the fractional D-brane wraps the
cycle
\be
\label{fractional}
\Pi^F_a\, =\, {1\over 4}\, \Pi^B_a + {1\over 4} \sum_{I \in
  S_\theta^a} \epsilon^\theta_{a,I}\, \Pi^\theta_{I,\,a}+ {1\over
  4}\sum_{J\in S_{\theta'}^a} \epsilon^{\theta'}_{a,J}\,
\Pi^{\theta'}_{J,\,a} + {1\over 4} \sum_{K\in S_{\theta\theta'}^a}
\epsilon^{\theta\theta'}_{a,K}\, \Pi^{\theta\theta'}_{K,\,a} \ee
where the $1/4$ factor indicates that one needs four fractional branes
in order to get a bulk brane. Also
$\epsilon^\theta_{a,I},\,\epsilon^{\theta'}_{a,J},\,\epsilon^{\theta\theta'}_{a,K}\,=\,\pm
1$ define the charge of the fractional brane $a$ with respect to the
massless fields living at the various fixed points. In the next
section we consider only $\epsilon^g_{J}=1$, as this is enough to
illustrate our main point. However, more complicated situations can be
arranged. A longer discussion can be found in \cite{bcms} for the
factorisable case.

\subsection{Tadpoles and K-theory}

We now mod out this theory by the orientifold action $\Omega{\mathcal
  R}$, where $\Omega$ is the world sheet parity and ${\mathcal R}$
acts by
$$ {\mathcal R}: z^I \to \overline{z}^I \,.$$
This action introduces four types of O6-planes associated to the
actions $\Omega{\mathcal R}$ $\Omega{\mathcal R}\theta$,
$\Omega{\mathcal R}\theta'$, $\Omega{\mathcal R}\theta\theta'$. The
corresponding O-plane can be either a O6$^{(-,-)}$ with negative RR
charge and tension or an exotic O6$^{(+,+)}$ with positive RR charge
and tension. Consistency with
discrete torsion implies that we need to introduce an odd number of
exotic O6-planes \cite{bcms}. In the rest of the paper, we take a single exotic
plane associated to O$_{\Omega{\mathcal
    R}}$. 
    
Taking this into account, we can define the homology classes of the
cycles wrapped by the O6-planes as follows 
\be \label{ocycle}
\Pi_{O6} =
\Pi_{\Omega{\mathcal R}} +\Pi_{\Omega{\mathcal R}\theta}
+\Pi_{\Omega{\mathcal R}\theta'} + \Pi_{\Omega{\mathcal
    R}\theta\theta'} 
\ee 
where 
\bea 
&&\Pi_{\Omega{\mathcal R}} \sim -2
[ a^1]\times[ a^2]\times[ a^3], \qquad \qquad \Pi_{\Omega{\mathcal
    R}\theta} \sim -2 [ b^1]\times[
b^2]\times[ a^3],\nonumber \\
&&\Pi_{\Omega{\mathcal R}\theta'} \sim -2 [ a^1]\times[ b^2]\times[
b^3], \qquad \qquad \Pi_{\Omega{\mathcal R}\theta\theta'} \sim -2 [
b^1]\times[ a^2]\times[ b^3].
\label{eq:oparts}
\eea
For factorisable lattices, the $\sim$ signs in (\ref{eq:oparts}) are
 equality signs. For non-factorisable lattices, additional
factors of two appear, if they are needed to obtain closed cycles
\cite{ftz}.

\smallskip

In the rest of the paper, we consider only the {\bf AAA}
orientifold\footnote{We are using the notation introduced in
  \cite{Blumenhagen:1999ev}.} for factorisable compactifications and
the (related) {\bf CCC} \cite{ftz} setup for non-factorisable ones.
With our conventions for the wrapping numbers (\ref{3-cycle}), the tadpole condition %
\be 
\sum_a N_a(\Pi_a^F + \Pi_{a'}^F) = 4\Pi_{O6} 
\ee 
can be expressed as untwisted
 \be
\begin{array}{lll} \vspace*{.2cm} \sum_a N_a n_a^1 n_a^2 n_a^3 & = &
  -16, \\\vspace*{.2cm} \sum_a N_a m_a^1 m_a^2 n_a^3 & = & -16,
  \\\vspace*{.2cm} \sum_a N_a m_a^1 n_a^2 m_a^3 & = & -16,
  \\\vspace*{.2cm} \sum_a N_a n_a^1 m_a^2 m_a^3 & = & -16,
\end{array}
\label{RRtad1}
\ee
plus twisted
\be
\begin{array}{lll} \vspace*{.2cm} \sum_a N_a n_a^1
  \epsilon_{a,I}^{\theta'} & = & 0, \\ \vspace*{.2cm} \sum_a N_a n_a^2
  \epsilon_{a,J}^{\theta\theta'} & = & 0, \\ \vspace*{.2cm} \sum_a N_a
  n_a^3 \epsilon_{a,K}^{\theta} & = & 0,
\end{array}
\label{RRtad2}
\ee
tadpole constraints.  The minus sign on the r.h.s.\ of the first
equation in (\ref{RRtad1}) reflects the appearance of an exotic
O-plane in the case with discrete torsion \cite{bcms}. As explained in
\cite{ftz} the number of O-planes is reduced in non-factorisable
lattices. However, for some wrapping numbers one unit corresponds to a
half-cycle as they refer to cycles on the factorisable lattice. These
two effects cancel resulting in the universal expressions
(\ref{RRtad1}), (\ref{RRtad2}).  The lattice dependence arises due to
the fixed point structure.

The tadpole conditions ensure the cancellation of non-Abelian
anomalies. On top of that, one has to impose K-theory constraints
\cite{Uranga:2000xp}. As discussed for example in the appendix of
\cite{bcms} these imply that a probe SU(2) stack of branes must host
an even number of fundamentals of SU(2). Following their lead, we
impose the sufficient condition that all our stacks contain an even
number of branes.

\subsection{Spectrum }

The resulting spectrum can now be calculated, as has been done in
\cite{bcms}. We reproduce it here for completeness. Firstly, D6-branes
wrapping three-cycles not invariant under $\Omega{\mathcal R}$ give
rise to the gauge group $U(N_a)$.  If two such branes intersect at an
angle open strings stretched between them will have massless
excitations. These give rise to chiral multiplets transforming under
the product of the two gauge groups on the branes.  The resulting
massless spectrum is given in table \ref{tcs},
\begin{table}[h]
  \renewcommand{\arraystretch}{1.5}
  \begin{center}
    \begin{tabular}{|c|c|}
      \hline
      Representation  & Multiplicity \\
      \hline
      $\Yasymm_a$
      & ${1\over 2}\left(\Pi'_a\cdot \Pi_a
        +\Pi_{{\rm O}6}\cdot \Pi_a\right)$  \\
      $\Ysymm_a$
      & ${1\over 2}\left(\Pi'_a\cdot \Pi_a-\Pi_{{\rm O}6} \cdot \Pi_a\right)$   \\
      $(\antifund_a,\fund_b)$
      & $\Pi_a\cdot \Pi_{b}$   \\
      $(\fund_a,\fund_b)$
      & $\Pi'_a\cdot \Pi_{b}$
      \\
      \hline
    \end{tabular}
    \caption{Chiral spectrum for intersecting D6-branes \cite{bcms}.}
    \label{tcs}
  \end{center}
\end{table}
where also the situation that brane $D_a$ intersects with its
orientifold image $D_{a^\prime}$ is included. In the latter case there
is only one gauge group factor due to the orientifold identification.
Further, branes that are invariant under the orientifold action
$\Omega{\mathcal R}\Pi_a^F = \Pi_a^F$ do not yield a unitary group but
rather a simplectic group $USp(2N_a)$. In the ${\mathbb
  Z}_2\times{\mathbb Z}_2'$ orbifold, fractional branes invariant
under $\Omega{\mathcal R}$ are those placed on top of an exotic
O6$^{(+,+)}$ plane. In our choice, they sit on top of the
O$_{\Omega{\mathcal R}}$ plane (see \cite{bcms} for further details).
Finally we recall that no adjoint fields from an $aa$ sector arise for rigid branes.

\subsection{Supersymmetry}

Although intersecting brane models which break supersymmetry
explicitly are not necessarily inconsistent, they usually suffer from
instabilities.  In order to avoid that to happen, we focus on models
with residual ${\cal N}=1$ supersymmetry. This amounts to the
condition that the angles $\theta^I_a$ ($I=1,2,3)$ every brane $D_a$
forms with the horizontal coordinate axes in each complex plane have
to add up to zero \cite{Berkooz:1996km},
\begin{equation}
  \theta_1 + \theta_2 + \theta_3 = 0\,\,\, \mbox{mod}\,\,\, 2\pi .
  \label{eq:susy}
\end{equation}
Often metric moduli can be adjusted such that (\ref{eq:susy}) is
satisfied. For later use, we specify the metric of the compact space
$G_{ab}$ ($a,b=1, \ldots, 6$) to be diagonal\footnote{Off-diagonal
  components are projected out.} in the coordinate basis of the $x^a$
(with $z^I = x^{2I -1} + \mbox{i}\, x^{2I}$ being the complex
coordinates in (\ref{eq:orbiac})) and define
\begin{equation}
  U^I = \sqrt{\frac{G_{2I,2I}}{G_{2I -1,2I-1}}},\,\,\, I=1,2,3 .
\end{equation}
For the factorisable lattice the $U^I$ are the complex structure
moduli of the $T^2$ factors.

\section{Explicit models}\label{example}

In this section we consider a concrete model which serves to
illustrate the model building rules, as well as how the number of
families restriction can be implemented once rigid branes are
introduced.  We do this in detail in a simple non-factorisable lattice
which serves to demonstrate our main result. We then construct the
factorisable version of the same model, in order to show how three
family models arise in that case too. Besides the family requirement,
we also need to impose twisted and untwisted tadpole conditions as
well as supersymmetry to the models. These constraints  impose
strong conditions on the brane wrapping numbers.

\subsection{Non-factorisable lattice \label{sec:fixt}}

As a minimal non-factorisable example, consider a lattice $\{e_i\}$
where the third and fifth lattice vectors are given by
\begin{equation}
  e_3 =\left( 0,0, 1,0,-1,0\right) \,\,\, , \,\,\, e_5 = \left( 0,0,
    1,0,1,0\right) 
\end{equation}
and keep the rest in a factorisable form ({\bf AAA}
lattice). Employing the Lefschetz fixed point theorem one finds that
there are 8 $\theta$-fixed tori, 16 $\theta^\prime$-fixed tori and 8
$\theta\theta^\prime$ fixed tori. The 8 $\theta$-fixed tori are
(underlined entries can be permuted)
\begin{equation}
  \begin{array}{l l}
    {\displaystyle \left( 0,0,0,0,x,y\right)} , &{\displaystyle  \left(
        \uline{ \frac{1}{2}, 0}, 0, 
        \uline{0 \hspace*{-.1in}\begin{array}{c}\\  
          \end{array}}, 
        x,y\right) ,} \\[2.5ex]
    {\displaystyle \left( \underline{\frac{1}{2},\frac{1}{2}}, 0,
        \uline{0 \hspace*{-.1in}\begin{array}{c}\\  
          \end{array}},
        x,y\right)} , & {\displaystyle \left(
        \frac{1}{2},\frac{1}{2} , 0 ,\frac{1}{2}, x,y\right) .}
  \end{array}
  \label{eq:thetafixed}
\end{equation}
Here, $x$ and $y$ are compactified on a two dimensional lattice
generated by $\left( 2,0\right)$ and $\left( 0,1\right)$.
The 16 $\theta^\prime$-fixed tori are
\begin{equation}
  \begin{array}{l l l }
    {\displaystyle \left( x,y,0,0,0,0\right)} , &{\displaystyle  \left(
        x,y,0,\uline{\frac{1}{2}},0, 
        \uline{0 \hspace*{-.1in}\begin{array}{c}\\  
          \end{array}}, 
      \right)} ,&{\displaystyle \left(
        x,y,0,\frac{1}{2},0,\frac{1}{2}\right) ,} \\[2.5ex] 
    {\displaystyle \left( x,y,\frac{1}{2},0,-\frac{1}{2},0\right)} ,
    &{\displaystyle  \left( 
        x,y, \frac{1}{2}, \uline{\frac{1}{2}} , -\frac{1}{2}, 
        \uline{0 \hspace*{-.1in}\begin{array}{c}\\  
          \end{array}}, 
      \right)} , & {\displaystyle \left(
        x,y,\frac{1}{2},\frac{1}{2},-\frac{1}{2},\frac{1}{2}\right) ,} \\[2.5ex]
    {\displaystyle \left( x,y,\frac{1}{2},0,\frac{1}{2},0\right)} ,
    &{\displaystyle  \left( 
        x,y, \frac{1}{2}, \uline{\frac{1}{2} }, \frac{1}{2}, 
        \uline{0 \hspace*{-.1in}\begin{array}{c}\\  
          \end{array}}, 
      \right)} ,&{\displaystyle \left(
        x,y,\frac{1}{2},\frac{1}{2},\frac{1}{2},\frac{1}{2}\right) ,} \\[2.5ex] 
    {\displaystyle \left( x,y,1,0,0,0\right) },
    &{\displaystyle  \left( 
        x,y, 1, \uline{\frac{1}{2}} , 0, 
        \uline{0 \hspace*{-.1in}\begin{array}{c}\\  
          \end{array}}, 
      \right)} , & {\displaystyle \left(
        x,y,1,\frac{1}{2},0,\frac{1}{2}\right)} .
  \end{array}
\end{equation}
Now, the compactification lattice for $\left( x,y\right)$ is generated
by $\left( 1,0\right)$ and $\left( 0,1\right)$.
Finally the 8 $\theta\theta^\prime$-fixed tori are
\begin{equation}
  \begin{array}{l l}
    {\displaystyle \left( 0,0,x,y,0,0\right)} , &{\displaystyle  \left(
        \uline{ \frac{1}{2}, 0}, x,y,0, 
        \uline{0 \hspace*{-.1in}\begin{array}{c}\\  
          \end{array}} 
      \right) ,} \\[2.5ex]
    {\displaystyle \left( \underline{\frac{1}{2},\frac{1}{2}},x,y, 0,
        \uline{0 \hspace*{-.1in}\begin{array}{c}\\  
          \end{array}}
      \right)} , & {\displaystyle \left(
        \frac{1}{2},\frac{1}{2} ,x,y, 0 ,\frac{1}{2}\right) ,}
  \end{array}
\end{equation}
where the compactification lattice for $\left( x,y\right)$ is
generated by $\left( 2,0\right)$ and $\left( 0,1\right)$.

Let us now compute the intersection number between two rigid
D6-branes given a compactification lattice. To do this, remember first that we denote a
D6-brane by its bulk wrapping numbers as (\ref{3-cycle}) \cite{ftz}: 

\begin{equation} \label{eq:dgen} D6_a = \left( m_a^1 \left[a^1\right]
    + n_a ^1 \left[ b^1\right] \right) \times \left( m_a ^2\left[
      a^2\right] + n_a ^2 \left[ b^2\right]\right) \times \left( m_a
    ^3\left[ a^3\right] + n_a ^3 \left[ b^3\right]\right) ,
\end{equation}
%

\noindent where the one-cycles are listed in (\ref{eq:halfyc}), and $m^i_a$,
$n_a^i$ ($i=1,2,3$) are integers, we see that the cycle
(\ref{eq:dgen}) is closed on the compactification lattice if
\begin{equation}\label{eq:so12cond}
  n_a ^2 =\mbox{even}\,\,\,\,\,\,\mbox{and} \,\,\,\,\,\,
  n_a ^3 = \mbox{even}\,,
\end{equation}
otherwise the brane has to wrap the  corresponding cycle of the
factorisable lattice twice \cite{ftz}.

Now, the contribution from the bulk piece can be expressed as: 
\be 
[\Pi_a^B] \cdot [\Pi_b^B]
= 4\, [\Pi_a^{T^6}] \cdot [\Pi_b^{T^6}] = 2\prod_{i=1}^3
(n_a^im_b^i-m_a^in_b^i) 
\ee 
where we have used the results in \cite{ftz} to compute the intersection number 
$ [\Pi_a^{T^6}] \cdot [\Pi_b^{T^6}] $.
Adding the contribution from the twisted parts, using
(\ref{fractional1}) and (\ref{fractional}), we find that the general
expression for the intersection number between fractional branes in the
present lattice can be written as follows:

\begin{eqnarray} 
 I_{ab} & = & \frac{1}{8} \prod_i^3 (n^i_a m^i_b - m_a^i
n_b^i) +\frac{\delta^\theta_{ab}}{4} \left(\frac{n^3_a}{2} m^3_b -
  m_a^3 \frac{n_b^3}{2}\right) + \frac{\delta^{\theta'}_{ab}
}{4}\left(n^1_a m^1_b - m_a^1 n_b^1\right) + \nonumber \\   & &\hskip1cm +
\frac{\delta^{\theta\theta'}_{ab}}{4} \left(\frac{n^2_a}{2} m^2_b -
  m_a^2 \frac{n_b^2}{2}\right) \,,
\label{inter1}
\end{eqnarray}
where $\delta^g_{ab}$ denotes the number of common $g$-fixed points
between brane stacks $a$ and $b$.

Computing the net number of families\footnote{We consider the case
  that the colour group is a subgroup of the gauge symmetry on stack
  $a$.} $I_{ab} - I_{a'b}$, 
\bea\label{families1} && I_{ab} - I_{a'b} = -\frac{1}{4}\Big[ m_a^3
n_b^3 m_b^1 m_b^2 n_a^1 n_a^2 + m_a^1 n_b^1 m_b^2 m_b^3 n_a^2 n_a^3 +
m_a^1 n_b^1 m_a^2 m_a^3 n_b^2 n_b^3 +
\nonumber \\
&&\hskip4cm + m_a^2 n_b^2 m_b^1 m_b^3 n_a^1 n_a^3 + m_a^3
n_b^3\,\delta^{\theta}_{ab} + m_a^1 n_b^1\,\delta^{\theta'}_{ab} +
m_a^2 n_b^2\,\delta^{\theta\theta'}_{ab} \Big] \,,
 \eea
it can be seen that odd numbers can be easily obtained. Indeed, 
one can check that if two branes have less than four fixed points
in common in some sectors, that is $\delta_{ab}^g \ne (4,4,4)$,  
as well as requiring suitable $m^i$'s to be odd, it is possible to have  an odd number of 
families. We show later that a similar condition applies to the factorisable case.

Finally in order to fully compute the spectrum, we need the intersection between the O6-planes and the fractional branes. In the present compactification lattice, the
cycles wrapped by the O6-planes (\ref{ocycle}), (\ref{eq:oparts}) can be written as  \cite{ftz}
\bea \label{O6lattice}
&&\Pi_{O6} = 2\Big[ (-1,0)\times(1,0)\times(2,0) + (0,1)\times(0,-1)\times(2,0)  \nonumber \\ 
&&\hskip4cm +  (1,0)\times(0,1)\times(0,-1) +(0,1)\times(2,0)\times(0,-1) \Big]\,,
\eea 
where the sign in the  first contribution comes from the exotic O6-plane. Then the intersection between the O6-planes with the branes can be computed using the results in \cite{ftz}, and boils down to the following expression
\bea 
\Pi_{O6}\cdot\Pi_a^F = \Pi_{O6}\cdot\Pi_a^{T^6} = \sum_{Oj}\prod_i(n^i_{Oj}m^i_a - m^i_{Oj}n^i_a)\,,
\label{O6a}
\eea
where $n^i_{Oj}$ correspond to  `wrapping numbers' for the O6-planes (\ref{O6lattice}) and the sum is over the four types of O6-planes. 


\subsection{Three Family Pati-Salam Model}

We are now ready to construct a three family Pati-Salam model
using rigid as well as hidden semi-rigid and non-rigid
branes. As discussed in \cite{bcms}, sometimes rigid branes can
combine with other rigid branes  to form a bulk brane which can move
off the fixed points. Moduli in the adjoint of a gauge group reappear
when this happens. Thus, such a set of branes forms  a non-rigid
brane. Branes which can be combined into a bulk brane have the same
wrapping numbers and cancelling twisted charges (see eq.~(\ref{RRtad2})). We call branes with
the same wrapping numbers and cancelling twisted charges in one
twisted sector,  semi-rigid. These can combine and form a brane
which can move away from the fixed points only in some directions.

A reversed view of this definition
starts with a bulk brane.  If its location is invariant under the
orbifold, it can split into its four fractional pieces obtained by
separating $\theta$, $\theta'$, $\theta\theta'$ images and adding
contributions from collapsed cycles such that each piece forms a
closed cycle in the blown up orbifold (see eq.~(\ref{fractional}) and
\cite{Blumenhagen:2002wn}). Keeping all such fractional pieces results
in a set which we call non-rigid, while keeping only the images of one
${\mathbb Z}_2$ factor, yields a semi-rigid set.

\bigskip

Let us start by describing the model building strategy.  We have seen that
in order to get an odd number of families, it is necessary to have some
of the fixed points different from their maximum value, that is $\delta_{ab}^g\ne (4,4,4)$ (four being the maximum in each entry).
Therefore, in order to cancel twisted tadpoles at all fixed points, it
will be necessary to introduce additional branes, compared to the case
when all fixed points are shared between branes\footnote{An easy way
  to cancel twisted tadpoles is to consider only branes which share
  all four fixed points, that is $\delta_{ab}^g = (4,4,4)$. In such case, 
  it is enough to fix appropriately the
  values of the wrapping numbers $n^i$ such that no tadpoles are left
  uncancelled (see (\ref{RRtad2})). This trick was used in the four
  family models constructed in \cite{bcms}. } (that is, when $\delta_{ab}^g = (4,4,4)$).
Care will be taken such  that these extra branes do not introduce exotic chiral matter. 
   A priori one will attach them to the hidden sector.  However, in order to 
obtain massless GUT Higgs pairs in the spectrum, it will be necessary to
 recombine one stack of the additional branes with the stack
carrying initially the SU(2)$_R$ gauge symmetry factor of the
Pati-Salam group (see below). 
Finally, we will be interested in models which preserve ${\mathcal
  N}=1$ supersymmetry. This will constrain further the wrapping
numbers of the brane stacks and fix some closed string moduli.
    
\smallskip

More explicitly, consider first a set of three {\it rigid branes} $\{a_1,\,
a_2,\,a_3\}$, the (a priori) visible sector, which share some, but not
all, fixed points in some sectors. In general this leads to some
uncancelled twisted tadpoles among themselves. Therefore it
is necessary to introduce an (a priori) hidden set of branes,  such
that the twisted tadpoles are cancelled.  In order to minimise this,
the two branes $\{a_2,\,a_3\}$ that will give rise to the gauge groups
SU(2)$_{L, R}$ in the Pati-Salam model, are taken such that they share
exactly the same set of fixed points, i.e.~$\delta_{a_2a_3}^g = (4,4,4)$. Hence, each one will contribute
to the same kind of twisted tadpoles, and we choose them such that
these tadpoles are cancelled between them.  Thus we are left with
uncancelled twisted tadpoles only from the stack $\{a_1\}$.  In order to
cancel these, we introduce a set of stacks $\{b_i\}$, such that {\it all}
twisted tadpoles from sets $\{a,\,b\}$ are cancelled. The set of
branes in all stacks $\{b_i\}$ have the same wrapping numbers and twisted
charges with respect to one of the ${\mathbb Z}_2$ factors, so that twisted tadpoles are
cancelled among them. Thus they form a stack of {\it semi-rigid}
branes.  Cancellation of untwisted tadpoles can at last be achieved
by introducing suitable sets of hidden sector branes, without
introducing new contributions to the twisted tadpoles. 
In the model we construct below, two more of these sets $\{c,\,d\}$ will be needed. 
Stacks within each of the sets $\{c\}$ 
can combine into bulk branes and hence they form {\it non-rigid} stacks.
Indeed, it is the requirement of unbroken residual supersymmetry which
restricts us to consider {\it all} hidden sector branes to be  semi-rigid or non-rigid.

Taking into account  all the requirements listed above, we end up with the semi  
realistic Pati-Salam-like  model specified in table \ref{tab:simp}. We perform in detail its analysis in what follows.

\begin{table}[h]
  \renewcommand{\arraystretch}{1.5}
  \begin{center}
    \begin{tabular}{|c||c|c|c|}
      \hline
      $N_\alpha $  &  $(n_\alpha ^{1},m_\alpha ^{1})$  &  $(n_\alpha
      ^{2},m_\alpha ^{2})$ 
      &  $(n_\alpha ^{3},m_\alpha ^{3})$ \\
      \hline
      \hline $N_{a_1} = 4$ & $(0,1)$ & $(0,-1)$ & $(2,0)$ \\
      \hline $N_{a_2} = 2$ & $(-1,1)$ & $(4,-3)$ & $(0,-1)$ \\
      \hline $N_{a_3} = 2$ & $(1,-3)$ & $(-4,1)$ & $(0,-1)$ \\
      \hline
      \hline $N_{b_1}= 2$ & $(-4,-1)$ &  $ (-4,-1)$  & $(-2,1)$ \\
      \hline $N_{b_2}= 2$ & $(4,1)$ &  $ (4,1)$  & $(-2,1)$ \\
      \hline
      \hline $N_{c_1}= 14$ & $(1,0)$ &  $ (1,0)$  & $(2,0)$ \\
      \hline $N_{c_2}= 14$ & $(-1,0)$ &  $ (-1,0)$  & $(2,0)$ \\
      \hline $N_{c_3}= 14$ & $(1,0)$ &  $ (-1,0)$  & $(-2,0)$ \\
      \hline $N_{c_4}= 14$ & $(-1,0)$ &  $ (1,0)$  & $(-2,0)$ \\
      \hline
      \hline $N_{d_1}= 12$ & $(1,0)$ &  $ (0,1)$  & $(0,-1)$ \\
      \hline $N_{d_2}= 12$ & $(-1,0)$ &  $ (0,1)$  & $(0,1)$ \\
      \hline
    \end{tabular}
    \caption{\small Wrapping numbers for the three family
      non-factorisable Pati-Salam model. \label{tab:simp}}
  \end{center}
\end{table}
%
   

\bigskip

Let us start by identifying the fixed points through which the visible and
hidden sector branes pass. These are explicitly listed  in table \ref{fxpoints}.
Next, we construct the basis of the twisted three-cycles as defined in
(\ref{twistedbase}). For brane $\{a_1\}$, the basis is given by
\bea 
&& [\alpha^\theta_{I_{a1},\,n}] = 2\, [E^\theta_{I_{a1}}]\otimes
[0,0,0,0,2,0]\,, \\
&& [\alpha^{\theta'}_{I_{a1},\,m}] = 2\, [0,1,0,0,0,0] \otimes
[E^{\theta'}_{I_{a1}}]\,,\\
&&[\alpha^{\theta\theta'}_{I_{a1},\,m}] = 2\, [0,0,0,1,0,0] \otimes
[E^{\theta\theta'}_{I_{a1}}]\,, 
\eea 
where $[E^g_{I_{a1}}]$\footnote{We are being sloppy here, using  the same symbol to denote fixed points, tori or cycles. However, it should be clear from the context what we are referring to.} correspond to the 4 fixed points associated to brane $\{a_1\}$ in each sector. These
are listed in the first column of table \ref{fxpoints}. From this
basis, we can construct the twisted 3-cycle which the brane wraps,
using (\ref{3cycle}):
\be [\Pi^\theta_{I,\,a_1}]= 1\cdot [\alpha^\theta_{I_{a1},\,n}] \,,
\qquad\qquad [\Pi^{\theta'}_{I,\,a_1}]= 1\cdot
[\alpha^{\theta'}_{I_{a1},\,m}] \,, \qquad\qquad
[\Pi^{\theta\theta'}_{I,\,a_1}]= -1\cdot
[\alpha^{\theta\theta'}_{I_{a1},\,m}] \,.  \ee
Finally, the full fractional cycle (\ref{fractional}), which the stack
$\{a_1\}$ wraps is given by
\be 
\Pi^F_{a_1}\, =\, {1\over 4}\, \Pi^B_{a_1} + {1\over 4} \sum_{I}^4
\, \Pi^\theta_{I,\,a_1}+ {1\over 4}\sum_{I}^4\,
\Pi^{\theta'}_{I,\,a_1} + {1\over 4} \sum_{I}^4\,
\Pi^{\theta\theta'}_{I,\,a_1} . 
\ee
For stacks $\{a_i\}$ ($i=2,3$), we have instead 
\bea
&& [\alpha^\theta_{I_{a_i},\,m}] = 2\, [E^\theta_{I_{a_i}}]\otimes  [0,0,0,0,0,1]\,, \\
&& [\alpha^{\theta'}_{I_{a_i},\,n}] = 2\, [1,0,0,0,0,0] \otimes
[E^{\theta'}_{I_{a_i}}]\,,
\quad \qquad [\alpha^{\theta'}_{I_{a_i},\,m}] = 2\, [0,1,0,0,0,0] \otimes [E^{\theta'}_{I_{a_i}}]\,,\\
&&[\alpha^{\theta\theta'}_{I_{a_i},\,n}] = 2\, [0,0,2,0,0,0] \otimes
[E^{\theta\theta'}_{I_{a_i}}]\,, \qquad \quad
[\alpha^{\theta\theta'}_{I_{a_i},\,m}] = 2\, [0,0,0,1,0,0] \otimes
[E^{\theta\theta'}_{I_{a_i}}]\,, 
\eea 
where $[E^g_{I_{a_i}}]$ correspond
to the four fixed points associated to brane $\{a_i\}$ in each sector (see
table \ref{fxpoints}). The twisted 3-cycle which the brane $\{a_2\}$ wraps
(stack $\{a_3\}$ is very similar) is then:
\bea && [\Pi^\theta_{I,\,a_2}]= -1\cdot [\alpha^\theta_{I_{a2},\,n}]
\,, \nonumber \\
&& [\Pi^{\theta'}_{I,\,a_2}]= -1\cdot [\alpha^{\theta'}_{I_{a1},\,m}]
+ 1\cdot [\alpha^{\theta'}_{I_{a2},\,m}] \,, \nonumber
\\ 
&& [\Pi^{\theta\theta'}_{I,\,a_2}]= \,4 \cdot
[\alpha^{\theta\theta'}_{I_{a2},\,n}] - 1\cdot
[\alpha^{\theta\theta'}_{I_{a2},\,m}] \,.  \eea
Thus, the full fractional cycle (\ref{fractional}), which the stack
$\{a_2\}$ wraps is given by (again, stack $\{a_3\}$ is very similar)
\begin{equation} 
\Pi^F_{a_2}\, =\, {1\over 4}\, \Pi^B_{a_2} + {1\over 4} \sum_{I}^4
\, \Pi^\theta_{I,\,a_2}+ {1\over 4}\sum_{I}^4\,
\Pi^{\theta'}_{I,\,a_2} + {1\over 4} \sum_{I}^4\,
\Pi^{\theta\theta'}_{I,\,a_2} .
\end{equation}
For all other branes, one can find
the fractional cycles in a similar fashion.

\smallskip

\begin{table}[h]
  \renewcommand{\arraystretch}{1.5}
  \begin{center}
    \begin{tabular}{|c||c|c|c|c|c|}
      \hline
      $\theta$ sector &  $a_1$ &  $a_{2,3}$ &  $b_i$ & $c_i$ & $d_i$\\
      \hline
      \hline $E_1^\theta$ & (0,0,0,0) & (0,0,0,0) & (0,0,0,0) &
      (0,0,0,0) & (0,0,0,0)  \\
      \hline $E_2^\theta$ & (0,1/2,0,0) & (1/2,1/2,0,0) & (0,1/2,0,0)
      & (0,0,1,0)$^\star$ & (1/2,0,0,0) \\
      \hline $E_3^\theta$ & (0,0,0,1/2) & (0,0,0,1/2) & (0,0,0,1/2)
      & (1/2,0,0,0) & (0,0,0,1/2) \\
      \hline $E_4^\theta$ & (0,1/2,0,1/2) &  (1/2,1/2,0,1/2)  & (0,1/2,0,1/2) & (1/2,0,1,0)$^\star$ & (1/2,0,0,1/2) \\
      \hline
      $\theta'$ sector &  &  &   &  &   \\
      \hline $E_1^{\theta'}$ & (0,0,0,0) &   (0,0,0,0)  & (0,0,0,0)
      &   (0,0,0,0)  & (0,0,0,0)  \\
      \hline $E_2^{\theta'}$ & (0,1/2,0,0) &  (0,1/2,0,0)  & (0,1/2,0,0) 
      & (1,0,0,0)  & (0,1/2,0,0)\\
      \hline $E_3^{\theta'}$ & (1,0,0,0) &  (0,0,0,1/2)  & (1,0,0,1/2) 
      & (0,0,1,0)   & (0,0,0,1/2)\\
      \hline $E_4^{\theta'}$ & (1,1/2,0,0) &   (0,1/2,0,1/2)  & (1,1/2,0,1/2) & (1/2,0,1/2,0) &  (0,1/2,0,1/2)\\
      \hline
      $\theta\theta'$ sector &   &   &   &  &   \\
      \hline $E_1^{\theta\theta'}$ & (0,0,0,0)  &   (0,0,0,0)  & (0,0,0,0)    		& (0,0,0,0)  &   (0,0,0,0) \\
      \hline $E_2^{\theta\theta'}$ & (0,0,1,0)$^\star $ &  (1/2,1/2,0,0)  & (0,1/2,0,0) & (0,0,1,0)$^\star $  & (1/2,0,0,0) \\
      \hline $E_3^{\theta\theta'}$ & (0,1/2,0,0)  &   (0,0,0,1/2)  & (0,0,0,1/2) & (1/2,0,0,0)   & (0,0,0,1/2) \\
      \hline $E_4^{\theta\theta'}$ & (0,1/2,1,0)$^\star $ &  
      (1/2,1/2,0,1/2)  & (0,1/2,0,1/2) & (1/2,0,1,0)$^\star$  &
      (1/2,0,0,1/2)\\ 
      \hline
    \end{tabular}
    \caption{\small Fixed points for the non-factorisable branes in
      the Pati-Salam model of table \ref{tab:simp}. \label{fxpoints}}
  \end{center}
\end{table}

Before proceeding to calculate the spectrum, we need to clarify some
subtleties regarding the fixed points denoted with a $\star$ in table
\ref{fxpoints}. Consider for example the point $(0,0,1,0)^\star$ in
the $\theta\theta'$ sector of brane $\{a_1\}$ (see table \ref{fxpoints}).
Suppose it denoted the locus of a fixed torus, as in section \ref{sec:fixt}, then 
it would be equivalent to zero.  However, here we are looking at  the one-cycle (or
collapsed three-cycle) wrapped by the D-brane and it matters in which
direction the one-cycle extends.  Consider the full cycle $(0,0;
0,-x; 1,0)$ for brane $\{a_1\}$, this is equivalent to $(0,0; 1,-x; 0,0)$
and it is therefore shifted in the third direction. One has to take this
into account when counting the number of common fixed points between a
pair of branes. In computing the intersection number between two branes,  the shifted (second) version has to be used. If the brane extended along the third
direction instead,  the fixed cycle would indeed be equivalent to
the one located at the origin and contribute only once to the counting
of common fixed points.  

We are now ready to calculate the chiral
spectrum arising from the Pati-Salam stacks of branes
$\{a,\,\,b\}$ and the auxiliary branes $\{c,\,\,d\}$. 
For reasons mentioned already and to be discussed shortly, we
assign the visible sector to the set $\left\{ a_1, a_2,
  a_3, b_1 \right\}$. The
spectrum arising from open strings stretched between different branes
within this set is displayed in table \ref{tab:simp1a}, where we
removed anomalous U(1) factors from the gauge groups.

\begin{table}[h]
  \renewcommand{\arraystretch}{1.15}
  \begin{center}
    \begin{tabular}{|c|c|c|}
      \hline
      Sector  &   $SU(4) \times SU(2)_L \times SU(2)_1 \times SU(2)_2 $ 
      &  $SU(2)\times USp(28)^4 \times SU(12)^2 $  \\
      \hline
      $(a_1\, a_2)$  &  $3\times (\overline{4}, 2,1,1)$ & 
      $(1; 1,1,1,1; 1,1)$ \\
      $(a_1\, a_3)$  & $3\times (4,1, 2,1)$
      &\textquotesingle\textquotesingle \\ 
      $(a_2\, a_3)$  &$14\times (1,2, 2,1)$&
      \textquotesingle\textquotesingle\\ 
      $(a_2'\, a_2)$  & $14\times (1,1,1,1)$ &
      \textquotesingle\textquotesingle\\ 
      $(a_3'\, a_3)$  & $12 \times (1,1,1,1) + 2 \times (1,1,3,1) $
      &\textquotesingle\textquotesingle \\ 
$(a_1\, b_1)$ & $3 \times (4,1,1,2)$ & \textquotesingle\textquotesingle \\
$ \left(a_1 ^\prime \, b_1\right)$ & $3 \times (\overline{4},1,1,2)$ &
\textquotesingle\textquotesingle\\ 
$\left( a_2 \, b_1\right)$ & $23 \times (1,2,1,2)$ &
\textquotesingle\textquotesingle\\ 
$\left( a_3 \, b_1 \right)$ & $15 \times (1,1,2,2)$
&\textquotesingle\textquotesingle\\ 
$\left(b_1\, b_1^\prime\right)$ &$6 \times (1,1,1,3) + 16 \times
(1,1,1,1)$ &   \textquotesingle\textquotesingle\\        
\hline
    \end{tabular}
    \caption{Model of table \ref{tab:simp}: Massless spectrum from
      open strings stretching between 
      different branes within the `visible sector set' $\left\{ a,
        b_1\right\}$.   \label{tab:simp1a}}
  \end{center}
\end{table}
Notice further that the stack of branes $\{b\}$ has been arranged such that, not only the
$\{a_1\}$ twisted tadpoles are cancelled, but also such that the net
intersection between branes $\{b\}$ and $\{a_1\}$ vanishes. Second, chiral
matter arising from possible intersections between brane $\{a_1\}$ and
branes $\{c,\,d\}$ is eliminated by shifting the  latter branes away from the
origin, such that the twisted contribution, as well as the bulk parts
of the intersection numbers vanish (this possibility was also used in
the models of \cite{bcms}).  Thus, no extra chiral matter charged
under the Pati-Salam $U(4)$ arises.

Now let us look at  some of the  phenomenological implications of the model. As
far as the Standard Model matter and the electroweak Higgs is
concerned, it would have been enough to consider the branes of set
$\left\{a\right\}$ as the observable sector, and to identify the
SU(2)$_1$ with SU(2)$_R$ of the Pati-Salam model. However, the GUT
Higgs pair allowing to break the Pati-Salam group spontaneously to the
Standard Model group, would be missing. Attaching the stack $\{b_1\}$ 
to the visible sector yields a way to get GUT Higgs pairs as
well. Provided the potential is such that we can turn on vev's for 
bi-fundamentals of $SU(2)_1 \times SU(2)_2$ the product of the two
SU(2)'s can be broken to its diagonal subgroup. Identifying  that
diagonal subgroup with SU(2)$_R$ we obtain a Pati-Salam model with
 three generations of quarks and leptons as well as providing pairs
of electroweak and GUT Higgses. In our example model, there will be a surplus of
Higgs pairs of both types. 

If this mechanism is realised, we arrive at an interesting
conclusion. The requirements of obtaining three generations for quarks
and leptons as well as the presence of GUT Higgs pairs in the massless
spectrum are connected. To obtain three generations we had to leave
some of the twisted tadpoles arising from branes hosting standard model
matter uncancelled. The extra branes needed for twisted tadpole
cancellation now also contribute the GUT Higgs pair to the spectrum. 
 Choosing instead of the stack $\{b_1\}$ the
stack $\{b_2\}$ would give a very similar way of obtaining the GUT
Higgs pairs.   

Hence, the final gauge group arising from the visible sector is, as
shown in table \ref{tab:simp1a}. On the other hand, the hidden sector
yields the gauge groups U(2)$\times$ USp(28)$^2 \times$
U(12)$^2$. However, by taking some flat directions we can deform
these semi and non-rigid branes into bulk D-branes. Then the final gauge
group, upon eliminating anomalous U(1) factors is
SU(2)$\times$USp(28)$\times$SU(12).  

Finally, we look at supersymmetry. This imposes, from branes
$\{a_2,\,a_3\}$ the condition \be \arctan{U^1} +
\arctan{\frac{3\,U^2}{4}}=\frac{\pi}{2}\,; \qquad \quad
\arctan{3\,U^1} + \arctan{\frac{U^2}{4}}=\frac{\pi}{2}\,.  \ee
These two conditions provide the same relation between $U^1,\,U^2$,
namely:
\be\label{susy1} 
U^1 = \frac{4}{3\,U^2} \,.
\ee 
On the other hand,
supersymmetry on branes $\{b\}$ requires 
\be 
\arctan{\frac{U^1}{4}} +
\arctan{\frac{U^2}{4}}=\pi + \arctan{\frac{U^3}{2}}\,.  
\ee 
Plugging condition (\ref{susy1}) into this expression gives 
\be
\arctan{\frac{1}{3\,U^2}} + \arctan{\frac{U^2}{4}}=\pi +
\arctan{\frac{U^3}{2}}\,,
\ee 
which has a non trivial solution
\be\label{susy2} U^3 = \frac{8+6(U^2)^2}{11\,U^2} \,.
\ee
The other hidden branes $\{c,\,\,d\}$ do not give new constraints.

\bigskip

\subsection{The factorisable orbifold}

In this section we show that, following the same strategy as in the
previous section, it is possible to get a three family left-right
symmetric model from factorisable lattices with torsion, {\it without} the
need of introducing tilted tori, as in the case without torsion
\cite{csu,cll}.  

That this is the case, can be easily seen from the
analogue of (\ref{families1}) in the factorisable case. This is
simply:
\bea\label{families2} && I_{ab} - I_{a'b} = -\frac{1}{2}\Big[ m_a^3
n_b^3 m_b^1 m_b^2 n_a^1 n_a^2 + m_a^1 n_b^1 m_b^2 m_b^3 n_a^2 n_a^3 +
m_a^1 n_b^1 m_a^2 m_a^3 n_b^2 n_b^3 +
\nonumber \\
&&\hskip4cm + m_a^2 n_b^2 m_b^1 m_b^3 n_a^1 n_a^3 + m_a^3
n_b^3\,\delta^{\theta}_{ab} + m_a^1 n_b^1\,\delta^{\theta'}_{ab} +
m_a^2 n_b^2\,\delta^{\theta\theta'}_{ab} \Big]\,.
\eea
From this expression it becomes clear that once some of the
$\delta_{ab}^g$'s are taken different from its maximum value, i.e.~$\delta_{ab}^g \ne (4,4,4)$, one can
get odd numbers of families (again, combined with suitable choices of
the wrapping numbers). Moreover, it is also easy to see from this
expression why the models considered in \cite{bcms} gave always even
number of families.

As an explicit example, we consider the factorisable version of the
non-factorisable three family model discussed in the previous section.
The wrapping numbers and brane content are listed in table
\ref{factorisable}. 
\begin{table}[h]
  \renewcommand{\arraystretch}{1.5}
  \begin{center}
    \begin{tabular}{|c||c|c|c|}
      \hline
      $N_\alpha $  &  $(n_\alpha ^{1},m_\alpha ^{1})$  &  $(n_\alpha
      ^{2},m_\alpha ^{2})$ 
      &  $(n_\alpha ^{3},m_\alpha ^{3})$ \\
      \hline
      \hline $N_{a_1} = 4$ & $(0,1)$ & $(0,-1)$ & $(1,0)$ \\
      \hline $N_{a_2} = 2$ & $(-1,1)$ & $(4,-3)$ & $(0,-1)$ \\
      \hline $N_{a_3} = 2$ & $(1,-3)$ & $(-4,1)$ & $(0,-1)$ \\
      \hline
      \hline $N_{b_1}= 2$ & $(-4,-1)$ &  $ (-4,-1)$  & $(-1,1)$ \\
      \hline $N_{b_2}= 2$ & $(4,1)$ &  $ (4,1)$  & $(-1,1)$ \\
      \hline 
      \hline $N_{c_1}= 12$ & $(1,0)$ &  $ (1,0)$  & $(1,0)$ \\
      \hline $N_{c_2}= 12$ & $(-1,0)$ &  $ (-1,0)$  & $(1,0)$ \\
      \hline $N_{c_3}= 12$ & $(1,0)$ &  $ (-1,0)$  & $(-1,0)$ \\
      \hline $N_{c_4}= 12$ & $(-1,0)$ &  $ (1,0)$  & $(-1,0)$ \\
      \hline
      \hline $N_{d_1}= 4$ & $(0,1)$ &  $ (0,-1)$  & $(1,0)$ \\
      \hline $N_{d_2}= 4$ & $(0,1)$ &  $ (0,1)$  & $(-1,0)$ \\
      \hline
      \hline $N_{e_1}= 12$ & $(1,0)$ &  $ (0,1)$  & $(0,-1)$ \\
      \hline $N_{e_2}= 12$ & $(-1,0)$ &  $ (0,1)$  & $(0,1)$ \\
      \hline
    \end{tabular}
    \caption{\small Wrapping numbers for Pati-Salam model in the
      factorisable version of \ref{tab:simp}. \label{factorisable}}
  \end{center}
\end{table}
In this case, one can easily get the fixed point
structure, fractional cycles and intersection numbers using the
results of \cite{bcms}. For the spectrum, we simply show the
factorisable analogue of table \ref{tab:simp1a} in
table \ref{factspectruma}. 

\begin{table}[h]
  \renewcommand{\arraystretch}{1.15}
  \begin{center}
    \begin{tabular}{|c|c|c|}
      \hline
      Sector  &   $SU(4) \times SU(2)_L \times SU(2)_1 \times SU(2)_2 $ 
      &  $SU(2)\times USp(24)^4 \times SU(4)^2 \times SU(12)^2 $  \\
      \hline
      $(a_1\, a_2)$  &  $3\times (\overline{4}, 2,1,1)$ & 
      $(1; 1,1,1,1;1,1; 1,1)$ \\
      $(a_1\, a_3)$  & $3\times (4,1, 2,1)$
      &\textquotesingle\textquotesingle\\ 
      $(a_2\, a_3)$  &$26\times (1,2,
      2,1)$&\textquotesingle\textquotesingle\\ 
      $(a_2' \, a_2)$  & $6\times (1,3,1,1) + 20\times (1,1,1,1)$
      &\textquotesingle\textquotesingle\\ 
      $(a_3' \, a_3)$  & $14 \times (1,1,1,1) $
      &\textquotesingle\textquotesingle\\ 
$(a_1\, b_1)$ & $3 \times (4,1,1,2)$ & \textquotesingle\textquotesingle \\
$ \left(a_1 ^\prime \, b_1\right)$ & $3 \times (\overline{4},1,1,2)$
&\textquotesingle\textquotesingle \\ 
$\left( a_2 \, b_1\right)$ & $23 \times (1,2,1,2)$
&\textquotesingle\textquotesingle \\ 
$\left( a_3 \, b_1 \right)$ & $15 \times (1,1,2,2)$
&\textquotesingle\textquotesingle\\ 
$\left( b_1\,  b_1^\prime\right)$ &$18 \times (1,1,1,1) $
&\textquotesingle\textquotesingle \\        
\hline

    \end{tabular}
    \caption{Model of table \ref{factorisable}: Massless spectrum from
      open strings stretching between 
      different branes within the `visible sector set' $\left\{ a,
        b_1\right\}$.   \label{factspectruma}}
  \end{center}
\end{table}

Notice that, 
compared with the same type of model in the previous section, in the
factorisable case we need to introduce one extra stack of auxiliary
branes $\{e\}$,  in order to fully cancel untwisted tadpoles. In this respect,
the non-factorisable model is more attractive.


Notice also that, as in the previous section and in \cite{bcms},
intersections of the auxiliary branes with the $U(4)$ brane are
cancelled off by shifting those branes away from the
origin. Furthermore, the 
supersymmetry conditions (\ref{susy1}), (\ref{susy2}) are the same for 
this case.

\section{Discussion}

Motivated by the recent advances in intersecting D-brane model building, we
 studied $T^6/{\mathbb Z}_2\times {\mathbb Z'}_2$ orientifolds
in type IIA which admit rigid cycles and (non)-factorisable lattices.

We have shown that brane pairs which do not pass through the same set
of fixed points,
together with 
suitable choices of the wrapping numbers, allow for constructions of
three family non-factorisable models with semi realistic particle
spectra.  We demonstrated this explicitly in an ${\mathcal N}=1$, three
family, Pati-Salam example. 
There are no chiral exotics, Pati-Salam invariant mass terms for all
exotics are allowed. So, at the present stage, there are no obvious
reasons against the possibility that all exotic matter decouples. 
In addition to the requirement of three families and no chiral
exotics, tadpole cancellation 
and supersymmetry impose strong constraints on the wrapping numbers
for the brane configurations.  Hence one may expect only few models
with all these characteristics to be available.

A question which needs to be addressed  is to actually
check whether vector-like exotics can be decoupled. For that one needs
to analyse the superpotential in the effective theory as it arises
from the concrete intersecting brane model. However, we
emphasise that our original motivation was not to get a fully
realistic model at this stage, but to show how non-factorisable
lattices can give rise to three generation models.

Another interesting feature of the model we studied is that  the
same branes which are needed for cancelling the twisted tadpoles, $\{b
\}$, also produce the GUT Higgses needed to break the Pati-Salam group
down to the Standard Model group (these were not present in
\cite{bcms} if viewed as a four family model). Thus, these extra
branes are not just needed for twisted tadpole cancellation but also
for phenomenological reasons. However, as discussed in the text, the
mechanism requires non-zero vev's for some scalars. Again, it would be
desirable to turn on that vev under good knowledge of the
superpotential. (Since the corresponding multiplet is massless, it is
conceivable that there is indeed a flat direction along the required
vev.)

As a fortunate byproduct of our study of non factorisable lattices, we have found
that the very same strategy to get odd number of families works
equally well for the factorisable case without the need to introduce
tilted tori as it is necessary in the case of non-rigid D6-brane
models \cite{csu,cll}.  We showed this in the example of a factorisable version
of the Pati-Salam non-factorisable model presented. The matter and
gauge group content is very similar to the non-factorisable
case. However, the factorisable lattice requires the introduction of
one extra set of hidden branes, $\{e\}$, in order to fully satisfy
untwisted tadpole conditions. This in turn gives rise to a larger
gauge group as well as further extra matter.  In this respect, the
non-factorisable version of the Pati-Salam model we have studied is
favoured.
    
We expect  the same trick to get odd number of
families for other non-factorisable lattices to continue being valid,
upon appropriate choice of the wrapping numbers.  It is also plausible
that other non-factorisable lattices will require less number of
hidden branes in order to fully cancel tadpoles. Compared to our
minimal choice, however, the rank of the gauge group will be reduced
and it might become harder to embed the Standard Model gauge group.

We have just started exploration of these type of models, and thus our results
are far from  exhaustive. There are still several open problems that need
investigation. For example, we did not touch on the issue of
introducing fluxes along the lines of \cite{bcms}, to stabilise some
of the closed string moduli. Further, we concentrated on a
Pati-Salam model in order to sidestep the problem of imposing
K-theory constraints, which are automatically satisfied when the
number of branes per 
stack is even.  It would be important to explore possible strategies
to minimise the number of K-theory constraints such that three family
MSSM like models can be investigated (see for instance \cite{mss}).

In the case of heterotic compactifications on non-factorisable lattices
(for recent studies see
\cite{Faraggi:2006bs,Forste:2006wq,Takahashi:2007qc}) it has been
observed that the same massless spectra can be obtained from
factorisable orbifolds together with a generalised notion of discrete
torsion \cite{prrv}. If that observation is caused by some deeper
relation between 
generalised discrete torsion and non factorisable compactifications it
would be interesting to find a type II analogue. Such relations can
yield important input into landscape studies of type II
compactifications (for recent results see \cite{Gmeiner:2007zz} and
references therein).

\section*{Acknowledgements}
We thank Radu Tatar and Cristina Timirgaziu for collaboration at early stages
of this project. We are grateful to Steve Abel, Ralph Blumenhagen and
especially Fernando Marchesano for useful discussions.\\
I. Z. thanks Perimeter Institute for partial support and hospitality
while part of this work was done. She is also supported by an STFC
Postdoctoral Fellowship. This work was partially supported by the
European Union 6th framework program MRTN-CT-2004-503069
``Quest for unification'', MRTN-CT-2004-005104 ``ForcesUniverse'',
MRTN-CT-2006-035863 ``UniverseNet'' and
SFB-Transregio 33 ``The Dark Universe'' by Deutsche
Forschungsgemeinschaft (DFG).


\end{document}